\documentclass[prd,amsmath,amssymb,superscriptaddress,preprintnumbers,twocolumn,nofootinbib,10pt]{revtex4-1}%superscriptaddress,showpacs,

\pdfoutput=1

\usepackage{graphicx}
\usepackage{dcolumn}
\usepackage{bm}
\usepackage{amssymb}
\usepackage{latexsym}
\usepackage{booktabs}
\usepackage{amsmath}
\usepackage{multirow}
\usepackage{url}

\usepackage{float}
\usepackage[colorlinks=true, linkcolor=red, citecolor=blue]{hyperref}

\usepackage[normalem]{ulem}
\usepackage{color}
\usepackage{array}
\usepackage{enumerate}

\begin{document}

\title{A search for sterile neutrinos in interacting dark energy models using DESI baryon acoustic oscillations and DES supernovae data}

\author{Lu Feng}\thanks{These authors contributed equally to this paper.}
\affiliation{College of Physical Science and Technology, Shenyang Normal University, Shenyang 110034, China}
\affiliation{Key Laboratory of Cosmology and Astrophysics (Liaoning) \& College of Sciences, Northeastern University, Shenyang 110819, China}
\author{Tian-Nuo Li}\thanks{These authors contributed equally to this paper.}%\footnote{Lu Feng and Tian-Nuo Li contributed equally to this work}
\affiliation{Key Laboratory of Cosmology and Astrophysics (Liaoning) \& College of Sciences, Northeastern University, Shenyang 110819, China}
\author{Guo-Hong Du}
\affiliation{Key Laboratory of Cosmology and Astrophysics (Liaoning) \& College of Sciences, Northeastern University, Shenyang 110819, China}
\author{Jing-Fei Zhang}
\affiliation{Key Laboratory of Cosmology and Astrophysics (Liaoning) \& College of Sciences, Northeastern University, Shenyang 110819, China}
\author{Xin Zhang}\thanks{Corresponding author}
\email{zhangxin@mail.neu.edu.cn}
\affiliation{Key Laboratory of Cosmology and Astrophysics (Liaoning) \& College of Sciences, Northeastern University, Shenyang 110819, China}
\affiliation{Key Laboratory of Data Analytics and Optimization for Smart Industry (Ministry of Education), Northeastern University, Shenyang 110819, China}
\affiliation{National Frontiers Science Center for Industrial Intelligence and Systems Optimization, Northeastern University, Shenyang 110819, China}
%\affiliation{Key Laboratory of Data Analytics and Optimization
%for Smart Industry, (Northeastern University), Ministry of Education, Shenyang 110819, China}
%\affiliation{Center for High Energy Physics, Peking University, Beijing 100080, China}

\begin{abstract}
Sterile neutrinos can influence the evolution of the universe, and thus cosmological observations can be used to search for sterile neutrinos. In this study, we utilized the latest baryon acoustic oscillations data from DESI, combined with the cosmic microwave background data from Planck and the five-year supernova data from DES, to constrain the interacting dark energy (IDE) models involving both cases of massless and massive sterile neutrinos. We consider four typical forms of the interaction term $Q=\beta H \rho_{\rm de}$, $Q=\beta H \rho_{\rm c}$, $Q=\beta H_{0} \rho_{\rm de}$, and $Q=\beta H_{0} \rho_{\rm c}$, respectively. 
Our analysis indicates that the current data provide only a hint of the existence of massless sterile neutrinos (as dark radiation) at about the $1\sigma$ level. In contrast, no evidence supports the existence of massive sterile neutrinos. Furthermore, in IDE models, the inclusion of (massless/massive) sterile neutrinos has a negligible impact on the constraint of the coupling parameter $\beta$. The IDE model of $Q=\beta H \rho_{\rm c}$ with sterile neutrinos does not favor an interaction. However, the other three IDE models with sterile neutrinos support an interaction in which dark energy decays into dark matter.

\end{abstract}

\maketitle

\section{Introduction}

The standard cosmological model ($\Lambda$CDM), which incorporates a cosmological constant ($\Lambda$) and cold dark matter (CDM), has provided an excellent fit to cosmological observations since the discovery of cosmic acceleration \cite{SupernovaSearchTeam:1998fmf,SupernovaCosmologyProject:1998vns}.
However, on one hand, the $\Lambda$ in the $\Lambda$CDM model is plagued with several theoretical difficulties, such as the ``fine-tuning'' and ``cosmic coincidence'' problems~\cite{Carroll:2000fy,Weinberg:2000yb,Sahni:1999gb,Frieman:2008sn}.
On the other hand, as experimental sensitivity improves, tensions of varying significance emerge between different observations.
The most famous and persisting one is the Hubble constant $H_0$. The cosmic microwave background (CMB) observation from the Planck satellite mission~\cite{Planck:2018vyg} infers $H_0=67.36\pm0.54~{\rm km}~{\rm s}^{-1}~{\rm Mpc}^{-1}$, while the local distance ladder method gives $H_0=73.04\pm1.04~{\rm km}~{\rm s}^{-1}~{\rm Mpc}^{-1}$ \cite{Riess:2021jrx}, revealing a tension (known as the ``Hubble tension'') at more than $5\sigma$. Recently, the Hubble tension has been widely discussed in the literature (see, e.g., Refs.~\cite{Li:2013dha,Zhang:2014dxk,Yang:2018euj,Guo:2018ans,Verde:2019ivm,DiValentino:2019jae,Guo:2019dui,Hryczuk:2020jhi,Lin:2020jcb,Gao:2021xnk,Cai:2021wgv,DiValentino:2021izs,Vagnozzi:2021tjv,Vagnozzi:2021gjh,Kamionkowski:2022pkx,Gao:2022ahg,Vagnozzi:2023nrq,Lynch:2024hzh,Huang:2024erq}).
Therefore, these findings imply the necessity to explore new physics beyond the standard $\Lambda$CDM model to explain the accelerated expansion of the universe.

Among the various extensions of the $\Lambda$CDM model, interacting dark energy (IDE) models, in which it is considered that there is some direct, nongravitational coupling between dark energy (DE) and dark matter (DM), have been proposed and studied widely~\cite{Li:2009zs,Fu:2011ab,Szydlowski:2008by,Feng:2016djj,DiValentino:2019ffd,Zhang:2021yof,Wang:2021kxc,Nunes:2022bhn,Jin:2022tdf,Zhao:2022bpd,Forconi:2023hsj,Li:2023gtu,Yao:2023jau,Benisty:2024lmj,Halder:2024uao,Sabogal:2025mkp}. It has been shown that the IDE models can help alleviate some theoretical and observational problems, such as the cosmic coincidence problem~\cite{Comelli:2003cv,Cai:2004dk,Zhang:2005rg} and the Hubble tension~\cite{Yang:2018uae,Vagnozzi:2019ezj}. Therefore, the coupling between DE and DM is an important theoretical possibility, and it is rewarding to further test the IDE model using cosmological observations.

Additionally, considering sterile neutrinos in cosmological models can also help alleviate the Hubble tension to some extent~\cite{Zhang:2014dxk,Zhang:2015rha,Feng:2017nss,Zhao:2017urm,Feng:2017mfs,Feng:2017usu}.
The presence of anomalies in some short-baseline oscillation experiments~\cite{LSND:2001aii,Acero:2007su,Mueller:2011nm,Mention:2011rk,MiniBooNE:2013uba,Hayes:2013wra} has been explained by sterile neutrinos with mass around the eV scale.
However, other experiments~\cite{IceCube:2016rnb,DayaBay:2016lkk} did not detect such a signal, which casts doubt on the hypothesis of the existence of sterile neutrinos. Thus, sterile neutrinos remain an enigma in neutrino physics. Since sterile neutrinos can have some effects on the evolution of the universe, cosmological observations can provide an independent way to search sterile neutrinos. Further studies on sterile neutrinos in cosmology can be found in Refs.~\cite{deHolanda:2010am,Palazzo:2013me,Hamann:2013iba,Wyman:2013lza,Battye:2013xqa,Dvorkin:2014lea,Ko:2014bka,Archidiacono:2014apa,Li:2014dja,Bergstrom:2014fqa,DayaBay:2014fct,Archidiacono:2014nda,Knee:2018rvj,Feng:2019jqa,DiValentino:2021rjj,Feng:2021ipq,Chernikov:2022mdn,Pan:2023frx,Feng:2024mfx}.

Recently, the Dark Energy Spectroscopic Instrument (DESI) collaboration has presented new baryon acoustic oscillations (BAO) measurements~\cite{DESI:2024uvr,DESI:2024lzq} and new cosmological parameter constraint results~\cite{DESI:2024mwx}. The constraint on cosmological parameters was derived from BAO observations in galaxy, quasar and Lyman-$\alpha$ (Ly$\alpha$) forests during the first year of observations from DESI.
The DESI BAO data combined with the CMB data from Planck and Atacama Cosmology Telescope (ACT) and the supernova (SN) data from Dark Energy Survey Year 5 (DESY5), have provided evidence at the 3.9$\sigma$ level for the $w_0w_a$CDM model~\cite{DESI:2024mwx}.
Furthermore, the search for sterile neutrinos has been explored in studies of the $\Lambda$CDM, $w$CDM, and $w_0w_a$CDM models, and the results indicate that the latest observational data can influence the measurement of sterile neutrino parameters~\cite{Du:2025iow}.
This has prompted several studies aimed at constraining various aspects of cosmological physics using the latest observational data (such as the DESI BAO data); see, e.g. Refs.~\cite{Wang:2024hks,Cortes:2024lgw,Colgain:2024xqj,Wang:2024rjd,Qu:2024lpx,Wang:2024dka,Gomez-Valent:2024tdb,DiValentino:2024xsv,Yang:2024kdo,Chan-GyungPark:2024spk,Escamilla-Rivera:2024sae,Wang:2024pui,DESI:2024aqx,Tyagi:2024cqp,Giare:2024syw,Li:2024qso,Du:2024pai,Ye:2024ywg,Giare:2024gpk,Sabogal:2024yha,Escamilla:2024ahl,Reboucas:2024smm,Giare:2024ocw,Wang:2024tjd,Alestas:2024eic,Chan-GyungPark:2024brx,Specogna:2024euz,Wu:2024faw,Li:2024qus,Pang:2024wul,Jiang:2024viw,Wang:2024rus,Li:2025owk,Du:2025iow}.
In particular, the latest DESI BAO data combined with the Planck CMB and DESY5 SN data supports the existence of an interaction between DE and DM~\cite{Li:2024qso}.
Thus it is necessary to explore the impact of these combined datasets on sterile neutrino measurements within the IDE framework.

In this study, we revisit the constraints on sterile neutrino parameters in IDE models using the latest observational data, including Planck CMB, DESI BAO, and DESY5 SN measurements. Our aim is to investigate how the latest observational data influence the constraints on sterile neutrino parameters in IDE models. 

This work is organized as follows. In Sec.~\ref{sec2}, we introduce the method and data utilized in this work. In Sec.~\ref{sec3}, we give the constraint results and provide some relevant discussions. The conclusion is given in Sec.~\ref{sec4}.

\section{Method and Data}\label{sec2}

\subsection{Method}\label{sec2.1}
In this paper, we focus on a special type of IDE model, i.e., vacuum energy interacting with CDM (I$\Lambda$CDM, hereafter), where DE is the vacuum energy with $w=-1$.
In this model, the energy conservation equations for DE and CDM satisfy
\begin{align}
\dot{\rho}_{\rm de} +3H(1+w)\rho_{\rm de}= Q,\label{eq1}\\
\dot{\rho}_{\rm c} +3 H \rho_{\rm c}= -Q,\label{eq2}
\end{align}
where a dot denotes the derivative with respect to the cosmic time $t$, $\rho_{\rm de}$ and $\rho_{\rm c}$ represent the energy densities of DE and CDM, respectively, $H$ is the Hubble parameter, and $Q$ denotes the energy transfer rate.
Since we currently have no fundamental theory to determine the form of $Q$, in this work, we consider four typical phenomenological forms of $Q$, i.e., $Q=\beta H \rho_{\rm de}$, $Q=\beta H \rho_{\rm c}$, $Q=\beta H_{0} \rho_{\rm de}$, and $Q=\beta H_{0} \rho_{\rm c}$, where the appearance of $H_{0}$ is only for a dimensional consideration, and $\beta$ denotes the coupling parameter.
From Eqs.~(\ref{eq1}) and (\ref{eq2}), it is known that $\beta > 0$ means CDM decaying into DE, $\beta < 0$ means DE decaying into CDM, and $\beta = 0$ indicates no interaction between DE and CDM.
For convenience, we denote the I$\Lambda$CDM models with $Q=\beta H \rho_{\rm de}$, $Q=\beta H \rho_{\rm c}$, $Q=\beta H_{0} \rho_{\rm de}$, and $Q=\beta H_{0} \rho_{\rm c}$ as the I$\Lambda$CDM1, I$\Lambda$CDM2, I$\Lambda$CDM3, and I$\Lambda$CDM4 models, respectively.

For the I$\Lambda$CDM model, the base parameter vector is ${\bf P}=\{\Omega_{\rm b} h^2,~\Omega_{\rm c} h^2,~H_0,~\tau,~\ln (10^{10}A_{\rm s}),~n_{\rm s},~\beta\},$
where $\Omega_{\rm b} h^2$ and $\Omega_{\rm c} h^2$ are the present-day physical densities of baryon and CDM, respectively, $H_0$ is the Hubble constant, $\tau$ is the reionization optical depth, $A_{\rm s}$ is the amplitude of primordial scalar perturbation power spectrum, $n_{\rm s}$ is the power-law spectral index, and $\beta$ is the coupling parameter characterizing the interaction strength between DE and CDM. We take $H_0$ as a free parameter instead of $\theta_{\rm {MC}}$ which is commonly used, because it depends on a standard non-interacting background evolution.

In this study, we consider two cases of massless and massive sterile neutrinos in the I$\Lambda$CDM model.
When the sterile neutrinos are considered to be massless (as the dark radiation), an additional parameter, the effective number of relativistic species $N_{\rm eff}$, should be added to the model; this is called I$\Lambda$CDM+$N_{\rm eff}$ model in this paper. Thus, there are eight base parameters in the I$\Lambda$CDM+$N_{\rm eff}$ model.
When the sterile neutrinos are considered to be massive, two extra free parameters, $N_{\rm eff}$ and the effective sterile neutrinos mass $m_{\nu,{\rm sterile}}^{\rm eff}$ need to be added to the model; this case is called I$\Lambda$CDM+$N_{\rm eff}$+$m_{\nu,{\rm{sterile}}}^{\rm{eff}}$ model in this paper. Thus, there are nine base parameters in the I$\Lambda$CDM+$N_{\rm eff}$+$m_{\nu,{\rm{sterile}}}^{\rm{eff}}$ model. Note that, the true mass of thermally distributed sterile neutrinos is given by $m_{\rm sterile}^{\rm thermal}=(N_{\rm eff}-3.044)^{-3/4}m_{\nu,{\rm sterile}}^{\rm eff}$. To avoid a negative $m_{\rm sterile}^{\rm thermal}$, $N_{\rm eff}$ must be larger than 3.044 in the universe with sterile neutrinos. In this case, the total mass of active neutrinos is fixed at $\sum m_\nu=0.06$ eV.

It should also be mentioned that when we consider the case where DE interacts with DM, then the DE is not a true background, and in this case we need to consider the perturbations of DE.
In a conventional way~\cite{Zhao:2005vj,Valiviita:2008iv,He:2008si}, DE is treated as a nonadiabatic fluid and to calculate the DE pressure perturbation in terms of adiabatic sound speed and the rest frame sound speed, thus the large-scale instability will occasionally occur in the IDE cosmology. To avoid this problem, we handle the perturbations of DE by using the extended parametrized post-Friedmann (ePPF) approach~\cite{Li:2014eha,Li:2014cee} (for the original version of PPF, see Refs.~\cite{Hu:2008zd,Fang:2008sn}). This approach can successfully avoid the large-scale instability problem in the IDE cosmology. In this study, we employ the ePPF method to treat the cosmological perturbations. For more detail information about the ePPF for the IDE cosmology, see Refs.~\cite{Li:2015vla,Zhang:2017ize,Feng:2018yew}.

We use the modified  {\tt CAMB}~\cite{Lewis:1999bs} and the public Markov-chain Monte Carlo package {\tt CosmoMC}~\cite{Lewis:2002ah} with inclusion of ePPF module and obtain the posterior probability distributions of the parameters.

\subsection{Data}\label{sec2.2}
In this paper, we use the following datasets of cosmological observations.

{\it The CMB data}: We use the Planck TT, TE, EE spectra at $\ell\geq 30$, the low-$\ell$ temperature commander likelihood, and the low-$\ell$ SimAll EE likelihood from the Planck 2018 data release~\cite{Planck:2018vyg}. Furthermore, we use the Planck PR4 lensing likelihood from combination of the NPIPE PR4 Planck CMB lensing reconstruction~\cite{Carron:2022eyg}.

{\it The DESI data}: We use 12 DESI BAO measurements in $0.1\leq z\leq4.2$~\cite{DESI:2024uvr,DESI:2024lzq,DESI:2024mwx}.\footnote{The DESI BAO data provided in \url{https://data.desi.lbl.gov/doc/releases/}.}
This dataset in different redshift bins include bright galaxy sample (BGS), luminous red galaxy (LRG), emission line galaxy (ELG), quasars (QSO), and Ly$\alpha$ forest.

{\it The DESY5 data}: We use the supernovae (SNe) data from the full 5-year data release of Dark Energy Survey (DES). This dataset comprises 194 low-redshift SNe ($0.025 < z < 0.1$), complemented by 1635 photometrically-classified SNe ($0.1 < z < 1.3$)~\cite{DES:2024jxu}, bringing the total number of SNe to 1829.

Note that in the search for sterile neutrinos, measurements of the growth of structure play a significant role. For example, observations of weak gravitational lensing (WL) and redshift space distortions (RSD) have been used to search for sterile neutrinos~\cite{Feng:2017nss,Zhao:2017urm,Feng:2017mfs,Feng:2017usu,Feng:2019jqa,Feng:2021ipq,Du:2025iow}. 
However, it is believed that significant, uncontrolled systematics still currently remain in these measurements to some extent. 
Therefore, we do not use measurements of the growth of structure for combined constraints in this paper. Instead, we consider only conservative observational datasets from CMB, DESI, and DESY5 to constrain the I$\Lambda$CDM models with sterile neutrinos.
In what follows, we will report and discuss the fitting results of the  I$\Lambda$CDM models with sterile neutrinos in light of CMB+DESI+DESY5 data combination.

\section{Results and discussion}\label{sec3}
\begin{figure*}[!htp]
\includegraphics[scale=0.9]{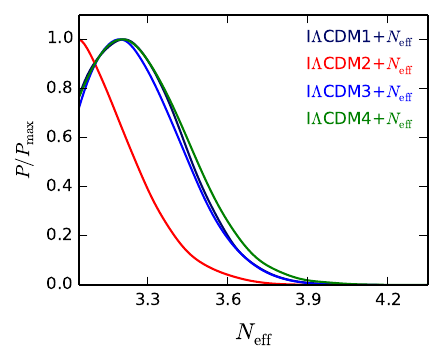}
\includegraphics[scale=0.9]{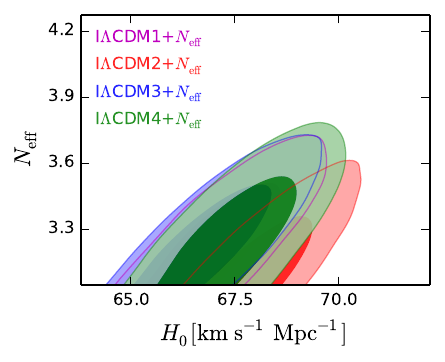}
\centering
\caption{\label{fig1} The one-dimensional marginalized posterior distributions for the $N_{\rm eff}$ (left) and two-dimensional marginalized posterior contours (1$\sigma$ and 2$\sigma$) in the $H_0$--$N_{\rm eff}$ plane (right) for the I$\Lambda$CDM+$N_{\rm eff}$ models using the CMB+DESI+DESY5 data.}
\end{figure*}

\begin{figure*}[!htp]
\includegraphics[scale=0.65]{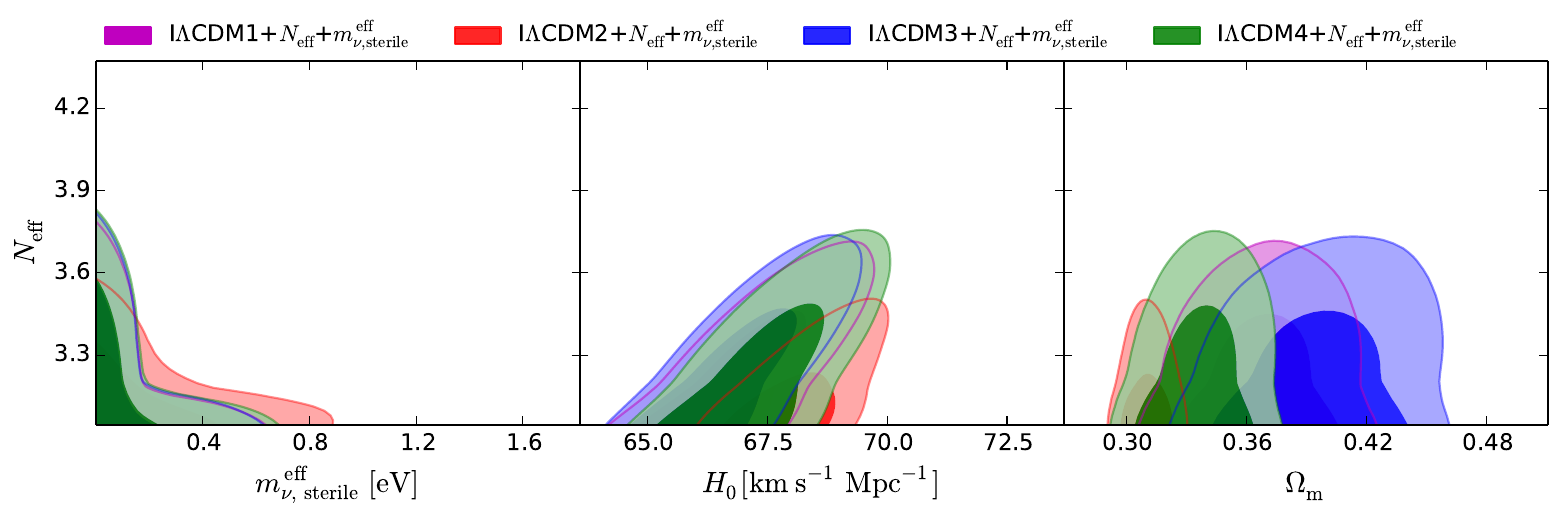}
\centering
\caption{\label{fig2} Two-dimensional marginalized posterior contours (1$\sigma$ and 2$\sigma$) in the $m_{\nu,{\rm{sterile}}}^{\rm{eff}}$-- $N_{\rm eff}$, $H_0$--$N_{\rm eff}$, and $\Omega_{\rm{m}}$--$N_{\rm eff}$ planes for the I$\Lambda$CDM+$N_{\rm eff}$+$m_{\nu,{\rm{sterile}}}^{\rm{eff}}$ models from the constraints of the CMB+DESI+DESY5 data combination.}
\end{figure*}

\begin{figure*}[!htbp]
\includegraphics[scale=0.9]{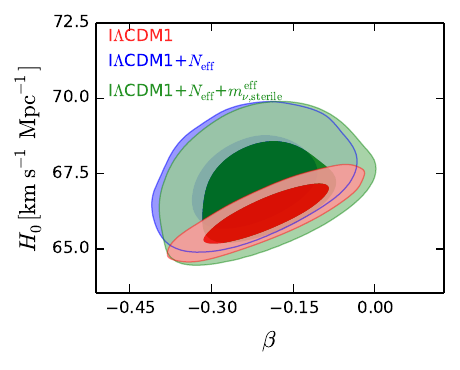}
\includegraphics[scale=0.9]{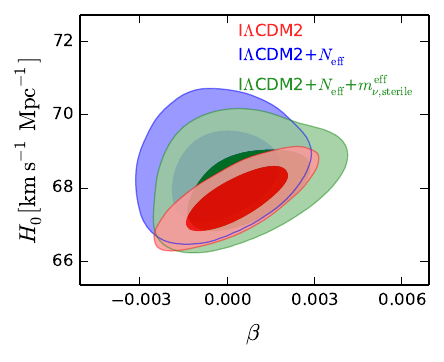}
\includegraphics[scale=0.9]{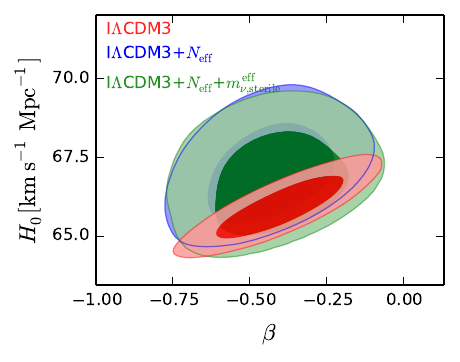}
\includegraphics[scale=0.9]{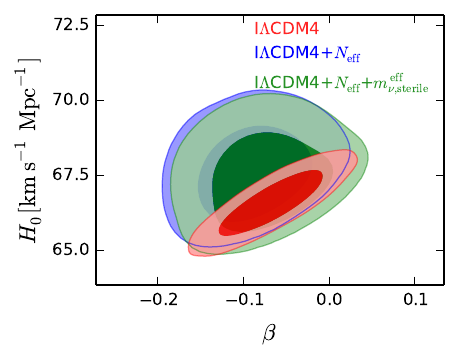}
\centering
\caption{\label{fig3}Two-dimensional marginalized posterior contours (1$\sigma$ and 2$\sigma$) in the $\beta$--$H_0$ plane for the I$\Lambda$CDM, I$\Lambda$CDM+$N_{\rm eff}$, and I$\Lambda$CDM+$N_{\rm eff}$+$m_{\nu,{\rm{sterile}}}^{\rm{eff}}$ models from the constraint of the CMB+DESI+DESY5 data combination.}
\end{figure*}

In this section, we report the constraint results of the I$\Lambda$CDM+$N_{\rm eff}$ and I$\Lambda$CDM+$N_{\rm eff}$+$m_{\nu,{\rm{sterile}}}^{\rm{eff}}$ models using the CMB+DESI+DESY5 data, and discuss the implications of the results. The constraint results are displayed in Tables~\ref{tabless}--\ref{tabms} as well as Figs.~\ref{fig1}--\ref{fig3}. For direct comparison, we reproduce the constraint results of the I$\Lambda$CDM models~\cite{Li:2024qso} from the CMB+DESI+DESY5 data combination in Table~\ref{tabilcdm}.

%%%%%%%%%%%%%%%%%%%%%%%%%%%%%%%%%%%%%%%%%%
\begin{table*}\small
%\begin{table*}\small
\setlength\tabcolsep{1.0pt}
\renewcommand{\arraystretch}{1.5}
\caption{\label{tabless}Fitting results of the I$\Lambda$CDM+$N_{\rm eff}$ models by using the CMB+DESI+DESY5 data combination. We quote $\pm 1\sigma$ errors for the parameters, but for the parameters that cannot be well constrained, we quote the $2\sigma$ upper limits. Here, $H_0$ is in units of ${\rm km}\ {\rm s^{-1}}\  {\rm Mpc^{-1}}$.}
\centering
\begin{tabular}{ccccccccccccccccccc}
\hline 
 && I$\Lambda$CDM1+$N_{\rm eff}$ &&  I$\Lambda$CDM2+$N_{\rm eff}$ && I$\Lambda$CDM3+$N_{\rm eff}$  && I$\Lambda$CDM4+$N_{\rm eff}$\\
\hline
$\Omega_{\rm{m}}$&&$0.371\pm0.020$&&$0.310\pm0.007$&&$0.398\pm0.025$&&$0.339\pm0.015$\\
$H_0$&&$67.32^{+0.87}_{-1.11}$&&$68.46^{+0.68}_{-0.94}$&&$67.09^{+0.87}_{-1.13}$&&$67.68^{+0.89}_{-1.13}$\\
$\beta$&&$-0.226\pm0.072$&&$-0.0002^{+0.0012}_{-0.0011}$&&$-0.440\pm0.130$&&$-0.089\pm0.040$\\
$N_{\rm eff}$&&$3.293^{+0.072}_{-0.242}$&&$<3.480$&&$3.290^{+0.070}_{-0.238}$&&$3.306^{+0.068}_{-0.260}$\\
\hline
\end{tabular}
%\end{table}
\end{table*}
%%%%%%%%%%%%%%%%%%%%%%%%%%%%%%%%%%%%%%%%%%

%%%%%%%%%%%%%%%%%%%%%%%%%%%%%%%%%%%%%%%%%%
\begin{table*}\small
%\begin{table*}\small
\setlength\tabcolsep{1.0pt}
\renewcommand{\arraystretch}{1.5}
\caption{\label{tabms}Fitting results of the I$\Lambda$CDM+$N_{\rm eff}$+$m_{\nu,{\rm{sterile}}}^{\rm{eff}}$ models by using the CMB+DESI+DESY5 data combination. We quote $\pm 1\sigma$ errors for the parameters, but for the parameters that cannot be well constrained, we quote the $2\sigma$ upper limits. Here, $H_0$ is in units of ${\rm km}\ {\rm s^{-1}}\  {\rm Mpc^{-1}}$ and $m_{\nu,{\rm{sterile}}}^{\rm{eff}}$ is in units of eV.}
\centering
\begin{tabular}{ccccccccccccccccccc}
\hline
&&I$\Lambda$CDM1+$N_{\rm eff}$+$m_{\nu,{\rm{sterile}}}^{\rm{eff}}$&&I$\Lambda$CDM2+$N_{\rm eff}$+$m_{\nu,{\rm{sterile}}}^{\rm{eff}}$&&I$\Lambda$CDM3+$N_{\rm eff}$+$m_{\nu,{\rm{sterile}}}^{\rm{eff}}$&&I$\Lambda$CDM4+$N_{\rm eff}$+$m_{\nu,{\rm{sterile}}}^{\rm{eff}}$\\
\hline
$\Omega_{\rm{m}}$&&$0.368\pm0.021$&&$0.310\pm0.007$&&$0.396\pm0.027$&&$0.337\pm0.016$\\
$H_0$&&$67.03^{+0.87}_{-1.23}$&&$68.12^{+0.59}_{-0.81}$&&$66.80^{+0.85}_{-1.24}$&&$67.44^{+0.89}_{-1.19}$\\
$\beta$&&$-0.206\pm0.076$&&$0.0006\pm0.0012$&&$-0.420^{+0.150}_{-0.130}$&&$-0.075^{+0.044}_{-0.043}$\\
$N_{\rm eff}$&&$<3.572$&&$<3.392$&&$<3.594$&&$<3.607$\\
$m_{\nu,{\rm{sterile}}}^{\rm{eff}}$&&$<0.375$&&$<0.604$&&$<0.381$&&$<0.400$\\
\hline
\end{tabular}
%\end{table}
\end{table*}
%%%%%%%%%%%%%%%%%%%%%%%%%%%%%%%%%%%%%%%%%%

%%%%%%%%%%%%%%%%%%%%%%%%%%%%%%%%%%%%%%%%%%
\begin{table*}\small
%\begin{table}\small
\setlength\tabcolsep{1.0pt}
\renewcommand{\arraystretch}{1.5}
\caption{\label{tabilcdm}Fitting results of the I$\Lambda$CDM models by using the CMB+DESI+DESY5 data combination. We quote $\pm 1\sigma$ errors for the parameters. Here, $H_0$ is in units of ${\rm km}\ {\rm s^{-1}}\  {\rm Mpc^{-1}}$.}
\centering
\begin{tabular}{ccccccccccccccccccc}
\hline
 &&  I$\Lambda$CDM1 && I$\Lambda$CDM2 &&  I$\Lambda$CDM3  &&  I$\Lambda$CDM4 \\
\hline
$\Omega_{\rm{m}}$&&$0.367\pm0.020$&&$0.310\pm0.007$&&$0.393\pm0.026$&&$0.335\pm0.015$\\
$H_0$&&$66.18\pm0.63$&&$67.72\pm0.56$&&$65.94\pm0.63$&&$66.59\pm0.69$\\
$\beta$&&$-0.197\pm0.071$&&$0.0003\pm0.0011$&&$-0.400\pm0.130$&&$-0.067\pm0.038$\\
\hline
\end{tabular}
%\end{table}
\end{table*}
%%%%%%%%%%%%%%%%%%%%%%%%%%%%%%%%%%%%%%%%%%

%%%%%%%%%%%%%%%%%%%%%%%%%%%%%%%%%%%%%%%%%%
%%%%%%%%%%%%%%%%%%%%%%%%%%%%%%%%%%%%%%%%%%

%%%%%%%%%%%%%%%%%%%%%%%%%%%%%%%%%%%%%%%%%%
%%%%%%%%%%%%%%%%%%%%%%%%%%%%%%%%%%%%%%%%%%

%%%%%%%%%%%%%%%%%%%%%%%%%%%%%%%%%%%%%%%%%%
%%%%%%%%%%%%%%%%%%%%%%%%%%%%%%%%%%%%%%%%%%

\subsection{The case of massless sterile neutrinos}
In the universe, the total relativistic energy density of radiation is given by
\begin{align}
\rho_r=[1+N_{\rm eff}\frac{7}{8}(\frac{4}{11})^{\frac{4}{3}}]\rho_\gamma,
\end{align}
where $\rho_\gamma$ is the photon energy density. The standard cosmological model has $N_{\rm eff}=3.044$~\cite{Akita:2020szl,Froustey:2020mcq,Bennett:2020zkv}. The detection of $\Delta N_{\rm eff}=N_{\rm eff}-3.044>0$ means the presence of extra relativistic particle species in the early universe, and in this study we take the fitting results of $\Delta N_{\rm eff}>0$ as evidence of the existence of massless sterile neutrinos.

In Table~\ref{tabless}, using the CMB+DESI+DESY5 data, we obtain $N_{\rm eff}=3.293^{+0.072}_{-0.242}$ for the I$\Lambda$CDM1+$N_{\rm eff}$ model, $N_{\rm eff}<3.480$ for the I$\Lambda$CDM2+$N_{\rm eff}$ model, $N_{\rm eff}=3.290^{+0.070}_{-0.238}$ for the I$\Lambda$CDM3+$N_{\rm eff}$ model, and $N_{\rm eff}=3.306^{+0.068}_{-0.260}$ for the I$\Lambda$CDM4+$N_{\rm eff}$ model, respectively. We find that, for I$\Lambda$CDM2+$N_{\rm eff}$, the $N_{\rm eff}$ cannot be well constrained using the CMB+DESI+DESY5 data. However, $N_{\rm eff}$ can be effectively constrained for I$\Lambda$CDM1+$N_{\rm eff}$,  I$\Lambda$CDM3+$N_{\rm eff}$, and I$\Lambda$CDM4+$N_{\rm eff}$ (see also Fig.~\ref{fig1}). The preference for $\Delta N_{\rm eff} > 0$ is at approximately the $1\sigma$ level for these three I$\Lambda$CDM+$N_{\rm eff}$ models using CMB+DESI+DESY5 data. Thus, we find that the conservative observation data combination can give a hint of the existence of massless sterile neutrinos (as dark radiation) at approximately $1\sigma$ level.

In fact, in previous studies~\cite{Feng:2017nss,Feng:2021ipq}, we also searched for massless sterile neutrinos in the $\Lambda$CDM, $w$CDM, and holographic dark energy (HDE) models using observational data available at that time. We found that conservative constraints from CMB, BAO, and SN data provide only an upper limit on $N_{\rm eff}$, but with the inclusion of large-scale structure (LSS) data, $N_{\rm eff}$ can be effectively constrained.
Furthermore, in Ref.~\cite{Feng:2024mfx}, we investigate the impact of future gravitational wave (GW) standard siren observations on constraints on $N_{\rm eff}$ in the $\Lambda$CDM model. We found that CMB+BAO+SN data provide only an upper limit of $N_{\rm eff}<3.464$. However, the addition of GW data (CMB+BAO+SN+GW) can significantly improve the constraint on $N_{\rm eff}$ and favors $\Delta N_{\rm eff} > 0$ at approximately the $3\sigma$ level. Thus LSS data and GW data play an important role in constraining the massless sterile neutrinos parameter $N_{\rm eff}$.

In this subsection, we search for massless sterile neutrinos using the latest conservative observational data (CMB+DESI+DESY5). Compared to previous conservative observations, we find that the latest conservative observation data significantly improve the constraint on $N_{\rm eff}$, with CMB+DESI+DESY5 favoring the existence of massless sterile neutrinos at approximately the $1\sigma$ level.

\subsection{The case of massive sterile neutrinos}
In Table~\ref{tabms}, using the CMB+DESI+DESY5 data, we obtain $N_{\rm eff}<3.572$ and $m_{\nu,{\rm{sterile}}}^{\rm{eff}}<0.375$ eV for the I$\Lambda$CDM1+$N_{\rm eff}$+$m_{\nu,{\rm{sterile}}}^{\rm{eff}}$ model, $N_{\rm eff}<3.392$ and $m_{\nu,{\rm{sterile}}}^{\rm{eff}}<0.604$ eV for the I$\Lambda$CDM2+$N_{\rm eff}$+$m_{\nu,{\rm{sterile}}}^{\rm{eff}}$ model, $N_{\rm eff}<3.594$ and $m_{\nu,{\rm{sterile}}}^{\rm{eff}}<0.381$ eV for the I$\Lambda$CDM3+$N_{\rm eff}$+$m_{\nu,{\rm{sterile}}}^{\rm{eff}}$ model, and $N_{\rm eff}<3.607$ and $m_{\nu,{\rm{sterile}}}^{\rm{eff}}<0.400$ eV for the I$\Lambda$CDM4+$N_{\rm eff}$+$m_{\nu,{\rm{sterile}}}^{\rm{eff}}$ model, respectively. We find that, under the constraint of CMB+DESI+DESY5 data, the fit results of $N_{\rm eff}$ and $m_{\nu,{\rm{sterile}}}^{\rm{eff}}$ cannot be well constrained for all the I$\Lambda$CDM+$N_{\rm eff}$+$m_{\nu,{\rm{sterile}}}^{\rm{eff}}$ models (see also Fig.~\ref{fig2}).
This result is consistent with previous findings~\cite{Du:2025iow}, which showed that the $\Lambda$CDM, $w$CDM, and $w_0w_a$CDM models provide upper limits on $N_{\rm eff}$ and $m_{\nu,{\rm{sterile}}}^{\rm{eff}}$ using the CMB (Planck and ACT)+DESI+DESY5 data combination. 

In Ref.~\cite{Du:2025iow}, we also found that when the LSS (1-year or 3-year weak-lensing data from DES) data is added to the CMB (Planck and ACT)+DESI+DESY5 data combination, for all dark energy models, $N_{\rm eff}$ can be effectively constrained and the fitting result of $m_{\nu,{\rm{sterile}}}^{\rm{eff}}$ also changes significantly. In particular, within the $w_0w_a$CDM model, we obtain $N_{\rm eff}=3.076^{+0.011}_{-0.017}$ and $m_{\nu,{\rm{sterile}}}^{\rm{eff}}=0.50^{+0.33}_{-0.27}$ eV, suggesting a non-zero sterile neutrino mass at approximately the $2\sigma$ significance level.

In our previous studies~\cite{Feng:2017nss,Feng:2017mfs,Feng:2021ipq}, we also searched for massive sterile neutrinos using observational data available at that time.
In Refs.~\cite{Feng:2017nss,Feng:2017mfs,Feng:2021ipq}, within the $\Lambda$CDM model, we found that conservative observational data provided upper limits on $N_{\rm eff}$ and $m_{\nu,{\rm{sterile}}}^{\rm{eff}}$. When considering dynamical dark energy models, such as $w$CDM and HDE~\cite{Feng:2017mfs,Feng:2021ipq}, we found that CMB+BAO+SN (or CMB+BAO) data still yielded upper limits on $N_{\rm eff}$ and $m_{\nu,{\rm{sterile}}}^{\rm{eff}}$. However, incorporating additional observational data, such as LSS, significantly improved these constraints. Moreover, in Refs.~\cite{Feng:2017mfs,Feng:2021ipq}, we also found that the nature of dark energy can significantly impact the measurement of $N_{\rm eff}$ and $m_{\nu,{\rm{sterile}}}^{\rm{eff}}$ (see also Ref.~\cite{Zhao:2017urm}).

In addition, several studies have investigated massive sterile neutrinos in the IDE models.
As shown in Refs.~\cite{Feng:2017usu,Feng:2019jqa}, searches for massive sterile neutrinos in the I$\Lambda$CDM, I$w$CDM, and IHDE models using previous observations, it was also found that, conservative observational data provides only upper limits on $N_{\rm eff}$ and $m_{\nu,{\rm{sterile}}}^{\rm{eff}}$. However, further incorporating LSS data improves constraints on these parameters.
Furthermore, in Ref.~\cite{Feng:2024mfx}, we also prospect how the future GW standard siren observations affect constraints on massive sterile neutrinos parameters in the $\Lambda$CDM model. We found that the CMB+BAO+SN data provides only an upper limit $m_{\nu,{\rm{sterile}}}^{\rm{eff}}<0.5789$ eV and $N_{\rm eff}<3.446$, but the addition of GW data can tighten the constraint on $m_{\nu,{\rm{sterile}}}^{\rm{eff}}$ significantly and improve the constraint on $N_{\rm eff}$, favoring $\Delta N_{\rm eff} > 0$ at more than the $1\sigma$ level.

These studies indicate that both previous and latest conservative observational data provide only upper limits on the sterile neutrino parameters $N_{\rm eff}$ and $m_{\nu,{\rm{sterile}}}^{\rm{eff}}$.
However, the inclusion of LSS or GW data significantly enhances the constraints on these parameters.

In this subsection, we investigate massive sterile neutrinos within the I$\Lambda$CDM model using the latest conservative observational data. We find that the CMB+DESI+DESY5 data combination provides only upper limits on the parameters $N_{\rm eff}$ and $m_{\nu,{\rm{sterile}}}^{\rm{eff}}$.  These results suggest that, in this case, no evidence supports the existence of massive sterile neutrinos.

\subsection{Cosmological parameters}
In this subsection, we wish to simply discuss the constraint results for the cosmological parameters, i.e., coupling parameter $\beta$ and Hubble constant $H_0$, in the I$\Lambda$CDM+$N_{\rm eff}$ and I$\Lambda$CDM+$N_{\rm eff}$+$m_{\nu,{\rm{sterile}}}^{\rm{eff}}$ models using the CMB+DESI+DESY5 data combination.

With the purpose of directly showing the impacts of (massless/massive) sterile neutrinos on constraints of the $\beta$, in this study, we perform an analysis for the I$\Lambda$CDM model (without sterile neutrinos), the I$\Lambda$CDM+$N_{\rm eff}$ model (with massless sterile neutrinos), and the I$\Lambda$CDM+$N_{\rm eff}$+$m_{\nu,{\rm{sterile}}}^{\rm{eff}}$ model (with massive sterile neutrinos), to make a comparison, and the detailed fit results are given in Tables~\ref{tabless}--\ref{tabilcdm} as well as Fig.~\ref{fig3}.

From the fitting results for $\beta$ presented in Tables~\ref{tabless}--\ref{tabms}, we find that, under the constraint of CMB+DESI+DESY5 data combination, the inclusion of (massless/massive) sterile neutrinos in the I$\Lambda$CDM2 model give the best constraints on the parameter $\beta$, and in this case $\beta=0$ is inside $1\sigma$ range, which indicates no interaction between DE and CDM. In other three I$\Lambda$CDM models (I$\Lambda$CDM1, I$\Lambda$CDM3, and I$\Lambda$CDM4) with (massless/massive) sterile neutrinos, the CMB+DESI+DESY5 data favors a negative coupling parameter $\beta$, and the deviations of $\beta$ from zero exceed the $2\sigma$ level. It is indicated that there is an interaction where DE decays into CDM. These results are basically consistent with the conclusion in previous study on the I$\Lambda$CDM models~\cite{Li:2024qso}. This also confirms the previous results in Ref.~\cite{Feng:2017usu}.

Therefore, we find that the consideration of (massless/massive) sterile neutrinos in the I$\Lambda$CDM models almost does not influence the constraint results of the coupling parameter $\beta$.

%Next, we examine whether $H_0$ tension can be effectively alleviated when (massless/massive) sterile neutrinos are considered in the I$\Lambda$CDM models.

In addition, sterile neutrinos can affect the constraints on $H_0$. From Tables~\ref{tabless}--\ref{tabilcdm}, we find that the central value of $H_0$ in the I$\Lambda$CDM models is lower. However, it increases slightly when (massless/massive) sterile neutrinos are included (see also Fig.~\ref{fig3}).

%However, this does not mean that the tension $H_0$ can be solved with this approach, as it introduces additional parameters to fit the observational data.

\section{Conclusion}\label{sec4}
The aim of this work is to search for sterile neutrinos in IDE models with energy transfer forms $Q=\beta H \rho_{\rm de}$, $Q=\beta H \rho_{\rm c}$, $Q=\beta H_{0} \rho_{\rm de}$, and $Q=\beta H_{0} \rho_{\rm c}$, respectively.
We consider the two cases of massless and massive sterile neutrinos, corresponding to the I$\Lambda$CDM+$N_{\rm eff}$ and I$\Lambda$CDM+$N_{\rm eff}$+$m_{\nu,{\rm{sterile}}}^{\rm{eff}}$ models, respectively. We utilize the latest observational datasets, including the BAO data from DESI, the SN data from DESY5, and the CMB data from Planck, to perform the analysis.

For the I$\Lambda$CDM+$N_{\rm eff}$ models, using the CMB+DESI+DESY5 data, $\Delta N_{\rm eff}>0$ is favored at approximately the $1\sigma$ level for the I$\Lambda$CDM1+$N_{\rm eff}$, I$\Lambda$CDM3+$N_{\rm eff}$, and  I$\Lambda$CDM4+$N_{\rm eff}$ models, respectively.
Therefore, the current CMB+DESI+DESY5 observational data give a hint of the existence of massless sterile neutrinos.
For the I$\Lambda$CDM+$N_{\rm eff}$+$m_{\nu,{\rm{sterile}}}^{\rm{eff}}$ models, only upper limits on both $N_{\rm eff}$ and $m_{\nu,{\rm{sterile}}}^{\rm{eff}}$ can be derived from the CMB+DESI+DESY5 data, indicating no evidence for the existence of massive sterile neutrinos.

Based on the constraints from CMB+DESI+DESY5, we find that for the I$\Lambda$CDM2 model with (massless/massive) sterile neutrinos, $\beta = 0$ is consistent with the CMB+DESI+DESY5 data within the $1\sigma$ range, which implies that this model is recovered to the $\Lambda$CDM universe. For the I$\Lambda$CDM1, I$\Lambda$CDM3, and I$\Lambda$CDM4 models with (massless/massive) sterile neutrinos, we find that negative values of $\beta$ are favored at more than the $2\sigma$ level using the CMB+DESI+DESY5 data, suggesting the presence of an interaction and a possible deviation from the standard $\Lambda$CDM model.
Moreover, we conclude that considering (massless/massive) sterile neutrinos does not significantly affect the constraint on $\beta$ by using the latest observational datasets.
In addition, we also find that incorporating (massless/massive) sterile neutrinos into I$\Lambda$CDM models slightly increases the fitted value of $H_0$. 

These findings underscore the significance of the latest conservative observational data combination in the search for sterile neutrinos within the framework of the IDE model. A natural next step is to search for sterile neutrinos by further considering measurements of the growth of structure, which will be examined further in future work.

\begin{acknowledgments}
This work was supported by the National Natural Science Foundation of China (Grant Nos. 12305069, 11947022, 12473001, 11975072, 11875102, and 11835009),
the National SKA Program of China (Grants Nos. 2022SKA0110200 and 2022SKA0110203),
and the Program of the Education Department of Liaoning Province (Grant No. JYTMS20231695).

\end{acknowledgments}

\bibliography{IDEsterile_DESI}

\end{document}